\documentclass[aps,pra,twocolumn,groupedaddress]{revtex4-1}
\usepackage{color}
\usepackage{dcolumn}
\usepackage{graphicx}
\usepackage{amsmath}
\usepackage{amssymb}
\usepackage{bm}

\begin{document}

%\preprint{}

\title{Critical dynamical properties of a first-order dissipative phase transition}


\author{W. Casteels$^1$}
\author{R. Fazio$^{2,3}$}
\author{C. Ciuti$^{1}$}
%\thanks{}
\affiliation{$^1$ Laboratoire Mat\'eriaux et Ph\'enom\`enes Quantiques, Universit\'e Paris Diderot, CNRS UMR 7162, Sorbonne Paris Cit\'e, 10 rue Alice Domon et Leonie Duquet 75013 Paris, France}
\affiliation{$^2$ ICTP, Strada Costiera 11, I-34151 Trieste, Italy}
\affiliation{$^3$ NEST, Scuola Normale Superiore and Istituto Nanoscienze-CNR, I-56126 Pisa, Italy}
%\no affiliation

\date{\today}

\begin{abstract}
We theoretically investigate the critical properties of a single driven-dissipative nonlinear photon mode. In a well defined thermodynamical limit of large excitation numbers,  the exact quantum solution describes a first-order phase transition in the regime where semiclassical theory predicts optical bistability. We study the behavior of the complex spectral gap associated to the Liouvillian superoperator of the corresponding master equation. We show that in this limit the Liouvillian gap vanishes exponentially and that the bimodality of the photon Wigner function disappears. The connection between the considered thermodynamical limit of large photon numbers for the single-mode cavity and the thermodynamical limit of many cavities for a driven-dissipative Bose-Hubbard system is discussed.

\end{abstract}

%\keywords{}

\maketitle

\section{Introduction}
In recent years the many-body physics of driven-dissipative optical systems has become a rapidly expanding research field (see for example Refs. \cite{ RevModPhys.85.299, ANDP:ANDP201200261, 0034-4885-79-9-096001, 2016arXiv160500383H, 2016arXiv160404433N}). This has led to an increasing interest in dissipative quantum phase transitions which have been theoretically studied for various systems such as coupled spins \cite{PhysRevA.86.012116, PhysRevA.93.023821, PhysRevX.6.031011, 2016arXiv160106857W} and dissipative Bose gases \cite{PhysRevB.94.085150,PhysRevB.79.165302, PhysRevX.4.021010, PhysRevLett.116.070407}. Currently the field is being lifted to the experimental realm. 
Examples are the spontaneous mirror-symmetry breaking in coupled photonic-crystal nanolasers \cite{2015NaPho...9..311H}, the observation \cite{2016arXiv160704892F} of the predicted photon-blockade breakdown phase transition\cite{PhysRevX.5.031028} , the report of bistability in one-dimensional circuit QED lattices \cite{2016arXiv160706895F} and the probing of the dynamic optical hysteresis in the quantum regime \cite{PhysRevA.93.033824, 2016arXiv160800260R}.

As in the case of quantum critical phenomena signatures of dissipative phase transitions should appear in the dynamical 
properties. The density-matrix $\hat{\rho}$ of an open quantum system is described by a linear master equation $\partial_t \hat{\rho} = \hat{\mathcal{L}} \hat{\rho}$ where $\hat{\mathcal{L}}$ is the so-called Liouvillian superoperator, having a spectrum of complex eigenvalues. The steady-state solution $\hat{\mathcal{L}} \hat{\rho} = 0$ corresponds to the zero eigenvalue of the Liouvillian. Dissipative phase transitions are expected to occur when the Liouvillian spectral gap \cite{PhysRevA.86.012116} closes in some thermodynamical limit. However, very little is known about the physical behavior of such a gap. To explore such uncharted territory, the study of paradigmatic and controlled model systems is of paramount importance for the fundamental understanding of dissipative phase transitions. A particular interesting class of systems to explore is the one represented by the driven-dissipative Kerr model, which describes a nonlinear optical resonator exhibiting optical bistability\cite{BONIFACIO1976172, PhysRevLett.40.1023, ROY1980133}.

In this paper, we explore the critical properties of the Liouvillian gap for a driven-dissipative (Kerr) nonlinear resonator. We show that, by considering a well defined thermodynamical limit of large excitation numbers, such a model describes a first-order phase transition. The thermodynamic limit is obtained by letting the nonlinearity going to $0$ and the driving intensity to $+ \infty$ while keeping constant their product. 
We determine the exponential vanishing of the complex Liouvillian gap and characterize its finite-size behavior. 
In this paper we show that a finite size scaling is crucial to determine the critical properties such as the critical driving strength in the thermodynamic limit.
As a perspective, we show that such thermodynamical limit of large excitation numbers for one single-mode resonator has a direct connection with the more standard limit of many sites in the driven-dissipative Bose-Hubbard model.

The paper is structured as follows: In Section \ref{model} a finite size scaling is introduced for the driven-dissipative Kerr model the steady-state properties are discussed. The critical power law behavior of the Liouvillian gap is then examined in Section \ref{LG}. In Section \ref{TL} the broader context of a 0D thermodynamic limit is discussed and as an intriguing perspective the link with the driven-dissipative Bose-Hubbard model is presented. Finally, in Section \label{Con}, the conclusions are drawn. 

\section{Finite-size scaling for the driven-dissipative Kerr model \label{model}}
We consider the following Hamiltonian for the driven Kerr model (with $\hbar = 1$):
\begin{eqnarray}
\hat{H} =  &&\omega_c \hat{a}^\dagger\hat{a} + \frac{U}{2} \hat{a}^\dagger\hat{a}^\dagger\hat{a}\hat{a} +\left(F e^{-i\omega_pt}\hat{a}^\dagger+\text{h.c.} \right),
\label{Ham}
\end{eqnarray}  
where $\hat{a}^\dagger$ ($\hat{a}$) creates (annihilates) an excitation in the resonator. The system parameters are $\omega_c$ for the cavity frequency, $U$ is the photon-photon interaction strength, and $F$ is the amplitude of the coherent drive with frequency $\omega_p$. Without loss of generality the drive amplitude $F$ will always be considered real. The cavity losses can be described within the Born-Markov approximation resulting in the following Lindblad master equation for the density matrix $\hat{\rho}$:
\begin{eqnarray}
\frac{\partial\hat{\rho}}{\partial t}= &&i\left[\hat{\rho},\hat{H}\right] + \frac{\gamma}{2}\left(2\hat{a} \hat{\rho}\hat{a}^\dagger  - \hat{a}^\dagger\hat{a} \hat{\rho}-\hat{\rho}\hat{a} ^\dagger\hat{a}  \right), 
\label{eq:Master}
\end{eqnarray}
with $\gamma$ the dissipation rate. From now on we will consider the frame rotating at the drive frequency $\omega_p$ which removes the time dependence of the Hamiltonian. The relevant parameter is then the frequency detuning $\Delta = \omega_p  - \omega_c$ between the drive and the cavity. 
%In general a linear master equation such as Eq. (\ref{eq:Master}) can be written as $\partial_t\hat{\rho} = \hat{\mathcal{L}}\hat{\rho}$ where $\hat{\mathcal{L}}$ is the Liouvillian superoperator. The spectrum of complex eigenvalues of the Liouvillian plays a crucial role. The steady-state corresponds to the zero eigenvalue. The Liouvillian gap $\lambda$ corresponds to the complex non-zero eigenvalue of $\hat{\mathcal{L}}$ whose real part is the closest to zero. The quantity $1/\vert \text{Re} (\lambda) \vert$ is the longest relaxation timescale of the system. In presence of a dissipative phase transition, the Liouvillian gap is expected to close at the critical point \cite{PhysRevA.86.012116}.

\begin{figure}[t!]
  \includegraphics[scale=0.45]{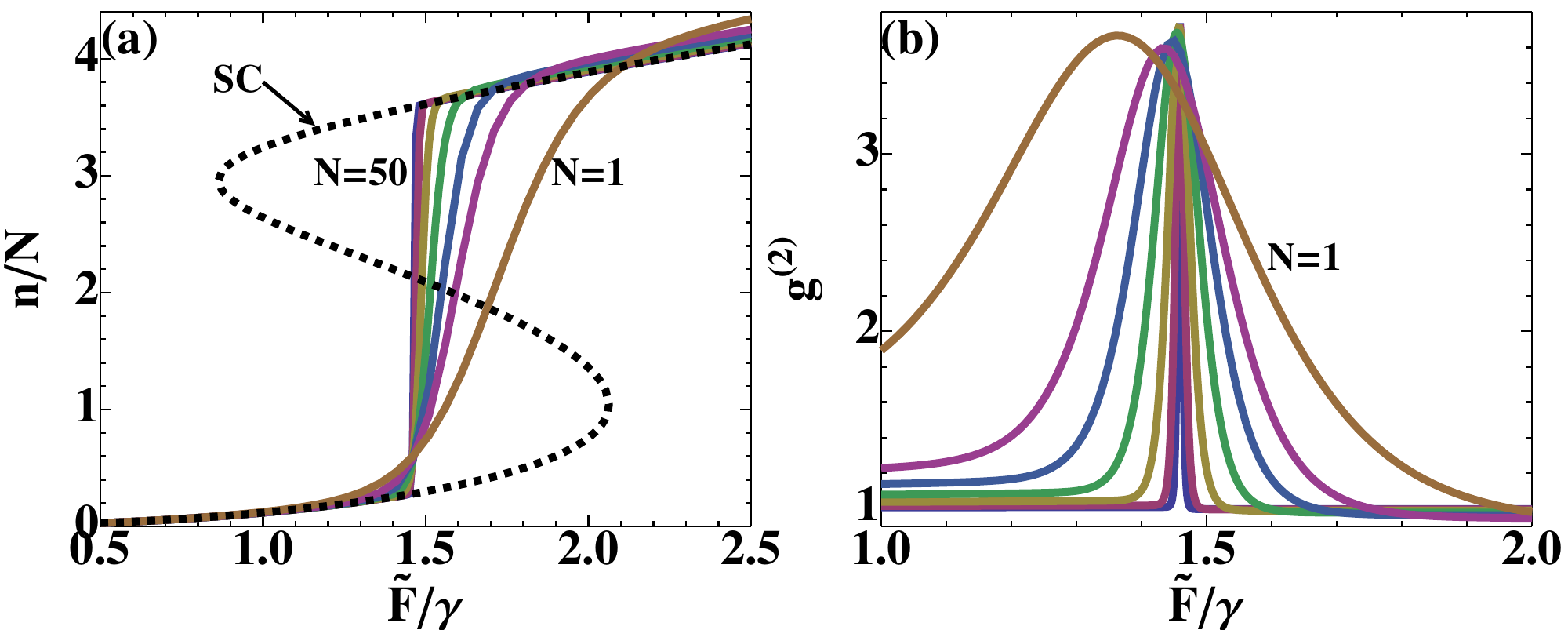}
  \caption{\label{Fig1} (a) The rescaled photon density $n/N$ (a) and (b) the second-order correlation function $g^{(2)}$ (b) as a function of the rescaled drive amplitude $\tilde{F}$ for $\Delta = 3\gamma$ and $\tilde{U} = \gamma$. The semiclassical prediction is also presented in (a) (SC, dashed line). Different curves correspond to $N = 1, 2, 3, 5, 10, 25$ and $50$.}
\end{figure}

We introduce the dimensionless parameter $N$ such that $U = \tilde{U}/N$ and $F = \sqrt{N}\tilde{F}$ and we will consider the limit $N \to \infty$. In such a limit $UF^2$ is constant and the number of excitations diverges (see later). This will be the thermodynamic limit we consider. 
In this limit quantum fluctuations become negligible and the system behaves (semi-)classically. The Lindblad master equation (\ref{eq:Master}) then reduces to an equation of motion for the coherent field amplitude $\alpha = \langle \hat{a}\rangle$ \cite{RevModPhys.85.299}:
\begin{eqnarray}
i\partial_t\tilde{\alpha} = \left(-\Delta  - i \frac{\gamma}{2} + \tilde{U} |\tilde{\alpha}|^2\right)\tilde{\alpha} +\tilde{F},
\label{GP}
\end{eqnarray}
with $\tilde{\alpha} = \alpha/\sqrt{N}$ the rescaled field amplitude. Eq. (\ref{GP}) is independent of $N$ and the photon number scales as $n = |\alpha|^2 \propto N$. This confirms that $N \rightarrow \infty$ corresponds to a well defined thermodynamic limit with an infinite number of photons. 
For a detuning $\Delta > \sqrt{3}/2\gamma$ there is a finite range of values for the drive amplitude for which the semiclassical Eq. (\ref{GP}) predicts three steady-state solutions with only two dynamically stable, typically denoted as optical bistability \cite{0305-4470-13-2-034} (see Fig. \ref{Fig1} (a)). 

We now use the analytical expressions derived in Refs. \cite{0305-4470-13-2-034, 1464-4266-1-2-005} for the steady-state properties corresponding to the master equation (\ref{eq:Master}) and examine the dependence on $N$. In Fig. \ref{Fig1} (a) and (b) the rescaled photon density $n/N$, with $n = \langle \hat{a}^{\dagger}\hat{a}\rangle$, and the normalised second-order correlation function $g^{(2)} = \langle \hat{a}^{\dagger}\hat{a}^{\dagger}\hat{a}\hat{a}\rangle/n^2$ are presented as a function of the rescaled drive amplitude $\tilde{F}$ for $\Delta = 3\gamma$, $\tilde{U} = \gamma$ and different values of $N$. In the limit $N \rightarrow \infty$ the density converges to one of the two stable semiclassical branches. Moreover, the transition between the two branches becomes increasingly sharp as $N$ is increased which suggests a discontinuous jump in the thermodynamic limit, as expected for a first order phase transition \cite{BONIFACIO1976172, PhysRevLett.40.1023}. The $g^{(2)}$-function in Fig. \ref{Fig1} (b) is strongly peaked around the transition which is due to the high fluctuations resulting from the switching between the two semiclassical branches. The width of the peak decreases as $N$ increases while the height is practically independent of $N$. 

\begin{figure}[t!]
  \includegraphics[scale=0.45]{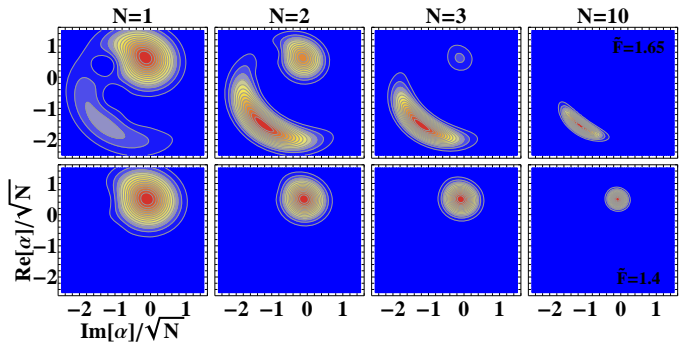}
  \caption{\label{Fig12} The photon Wigner function as a function of the real and imaginary part of the rescaled field $\alpha/\sqrt{N}$ for a drive amplitude $\tilde{F} = 1.65\gamma$ (upper row) and $\tilde{F} = 1.4\gamma$ (lower row) for $N = 1,2,3$ and $10$. Red corresponds to high values and green to zero (a different scale is used for the different panels). Other parameters are the same as in Fig. \ref{Fig1}. }
\end{figure}

In Fig. \ref{Fig12} the corresponding Wigner functions are presented for different values of $N$ and for two values of the rescaled drive amplitude $\tilde{F}$, one just above and one just below the transition. Below the transition a single peak is observed for all values of $N$. Above the transition and for $N$ sufficiently small the well-known bimodal shape typically attributed to optical bistability is observed \cite{PhysRevA.39.4675}. The system then explores two separate regions in phase space corresponding to the two semiclassical branches. This results in large fluctuations leading to the peak in $g^{(2)}$ in Fig. \ref{Fig1} (b). However, as $N$ is increased the relative weight of one of the peaks increases and for large $N$ only a single peak remains. This shows that in the thermodynamic limit the Wigner function consists of a single peak for all values of $F$. This is in agreement with the behavior of the density in Fig. \ref{Fig1} (a) which around the transition is an average of the two semiclassical results for small $N$ but for increasing $N$ converges to one of the semiclassical branches. This is also consistent with the presence of a first order phase transition in the thermodynamic limit.   

\section{Critical behavior of the Liouvillian gap \label{LG}}
The previous observations raise the question about the fate of the second stable semiclassical solution in the quantum formalism. To gain further insight we now consider the Liouvillian gap $\lambda$ of the Liouvillian superoperator $\hat{\mathcal{L}}$ associated to the master equation (\ref{eq:Master}). This is the generalisation to a dissipative context of the energy gap for a closed system. 
For a closed system at equilibrium the energy gap closes at a phase transition \cite{book:577981} and recently it was realised that similarly for a dissipative phase transition the complex Liouvillian gap $\lambda$ closes \cite{PhysRevA.86.012116}. 
In the following we examine the behavior of $\lambda$ which is obtained by numerically diagonalizing the Liouvillian superoperator $\hat{\mathcal{L}}$ in the Fock basis. The Liouvillian gap is the complex non-zero eigenvalue of $\hat{\mathcal{L}}$ whose real part is closest to zero. The quantity $-1/Re[\lambda]$ is the largest relaxation timescale of the system. Convergence of the results has been carefully checked by varying the cutoff number of photons.

\begin{figure}[t!]
  \includegraphics[scale=0.43]{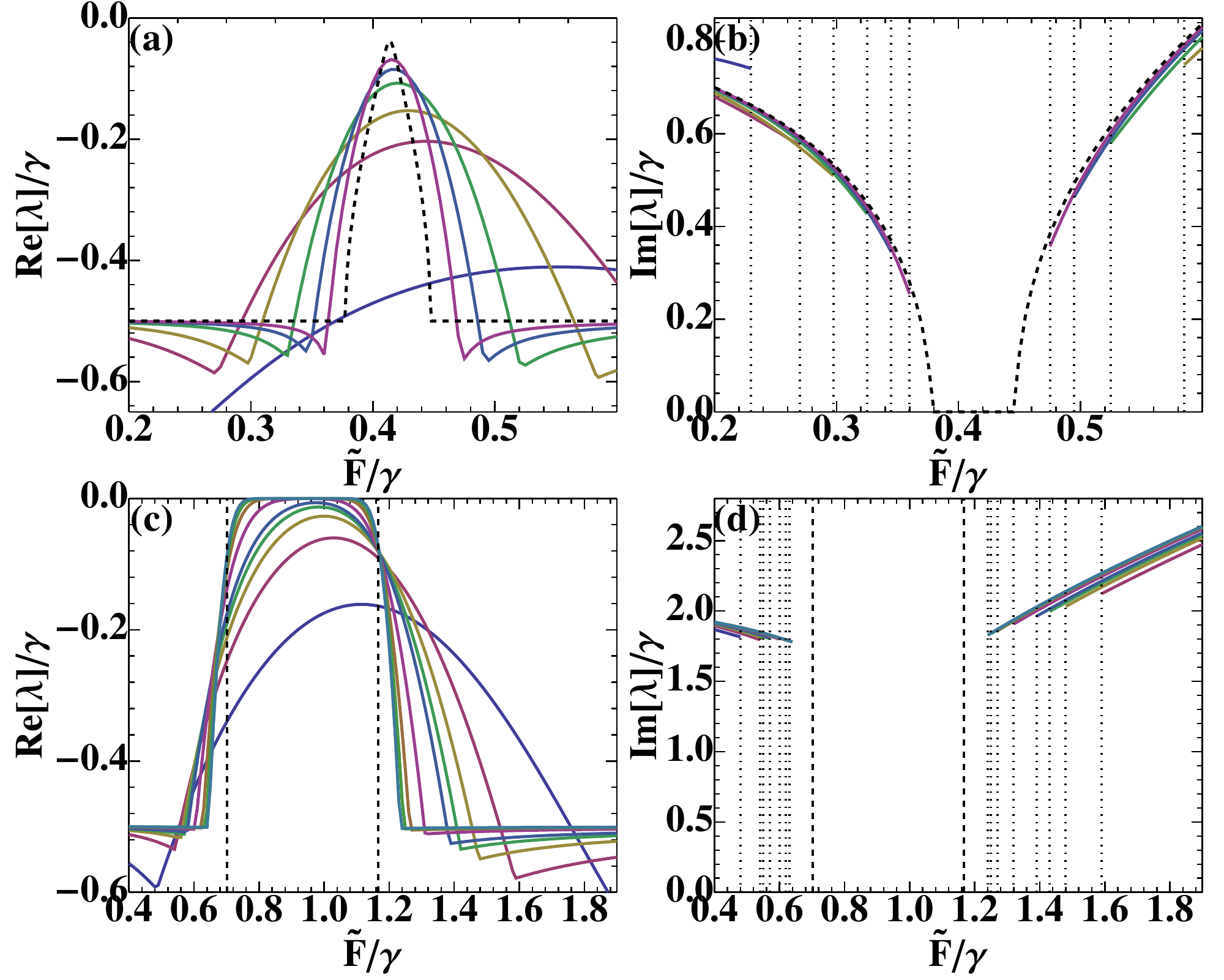}
  \caption{\label{Fig2} The real (a,c) and the imaginary (b,d) part of the Liouvillian gap $\lambda$ (units of $\gamma$) as a function of the rescaled drive amplitude $\tilde{F}$ (units of $\gamma$) for $\tilde{U}=\gamma$ and two values of the detuning: $\Delta = 0.8\gamma$ (a,b) and $\Delta = 2\gamma$ (c,d). The different curves correspond to different values $N = 1,5,10,25,50,100$ for $\Delta = 0.8\gamma$ and $N = 1,2,3,4,5,10,20,30,40$ for $\Delta = 2\gamma$ (as the curves approach the dashed lines $N$ increases). For the imaginary parts in (b) and (d) all curves drop to zero in a range whose size decreases with increasing $N$ (denoted by the dotted lines). In (a) and (b) the dashed lines correspond to the semiclassical linear response spectrum $\lambda_{LR}$. In (c) and (d) the dashed lines indicate the edge of the regime where the semiclassical approach predicts bistability. }
\end{figure}

In Fig. \ref{Fig2} (a) and (b) the real and the imaginary part of the Liouvilian gap are presented as a function of the rescaled drive amplitude $\tilde{F}$ for different values of $N$ with $\tilde{U} = \gamma$ and a detuning $\Delta = 0.8\gamma$ which is below the semiclassical threshold for bistability. We also present the real and imaginary part of the linearized spectrum $\lambda_{LR}$ around the steady-state semiclassical solution \cite{0305-4470-13-2-034, RevModPhys.85.299}. The imaginary part of $\lambda$ corresponds to an excitation frequency while $-Re[\lambda]$ is the corresponding damping rate. In Fig. \ref{Fig2} (a) and (b) we observe a region where the real part of $\lambda$ becomes suppressed and the imaginary part is equal to zero. The size of this region reduces as $N$ is increased and in the thermodynamic limit the results show that the Liouvillian gap $\lambda$ converges to the linear response spectrum $\lambda_{LR}$. From now on we will use the notation $\tilde{F}_c$ for the drive amplitude corresponding to the smallest value of $|Re[\lambda]|$.
In Fig. \ref{Fig2} (c) and (d) the real part and the imaginary part of the Liouvillian gap are presented as a function of the drive amplitude for different values of $N$, a rescaled nonlinearity $\tilde{U} = \gamma$ and a detuning $\Delta = 2\gamma$, for which the semiclassical approach predicts bistability. As before we find that around the transition there is a region where the imaginary part is zero and the absolute value of the real part is strongly suppressed. The size of this region reduces as $N$ is increased. In the same region the absolute value of the real part continues to decrease as $N$ is increased suggesting that the Liouvillian gap closes in the thermodynamic limit. 

\begin{figure}[t!]
  \includegraphics[scale = 0.4]{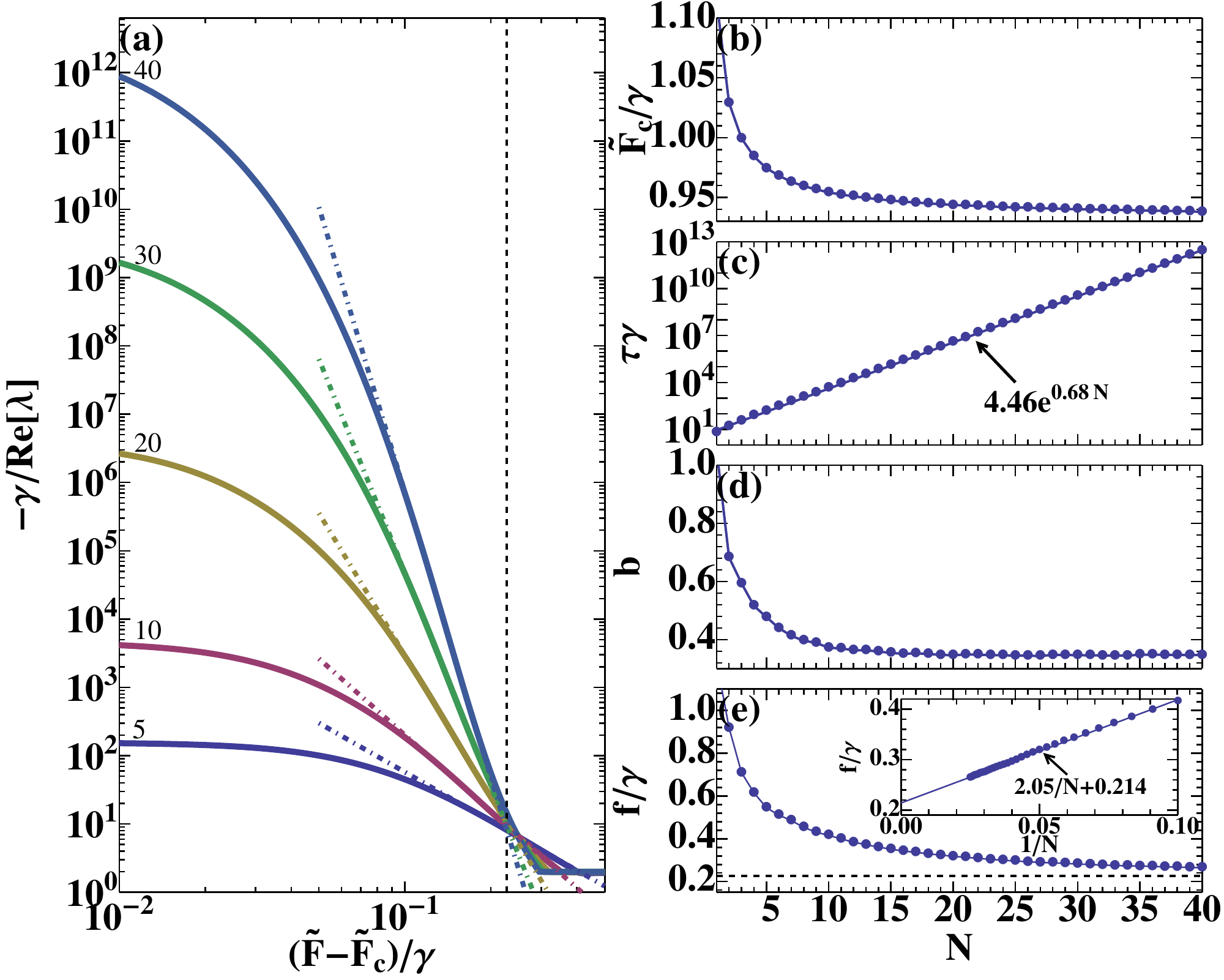}
  \caption{\label{Fig3} (a) The relaxation timescale $-1/Re[\lambda]$ (units of $\gamma^{-1}$) as a function of the distance from the transition point $\tilde{F} - \tilde{F}_c$ (units of $\gamma$) for $\tilde{U} = \gamma$ and $\Delta = 2\gamma$. The different curves correspond from top to bottom to $N = 40, 30, 20, 10$ and $5$. For the same parameters and as a function of $N$: (b) the drive amplitude $\tilde{F}_c$ at the transition point (units of $\gamma$); (c) the tunneling time $\tau$ (units of $\gamma^{-1}$); (d) the dimensionless fitting parameter $b$ and (e) the fitting parameter $f$ (units of $\gamma$). The dash-dotted lines in (a) are the power law fits (\ref{Fit}). The dashed lines in (a) and (e) indicate the drive amplitude where the semiclassical approach predicts the edge of the bistable region. The inset in (e) demonstrates the finite size scaling analysis that is used for the extrapolation to the thermodynamic limit. }
\end{figure}

In Fig. \ref{Fig3} (a) the relaxation timescale $-1/Re[\lambda]$ is presented as a function of $\tilde{F} - \tilde{F}_c$, for  $\tilde{U} = \gamma$, $\Delta = 2\gamma$ and different values of $N$. In Fig. \ref{Fig3} (b) we examine the dependence of the transition point $F_c$ on $N$. In the thermodynamic limit and for the considered parameters, $\lim_{N \to \infty} \tilde{F}_c \simeq 0.93\gamma$. At the transition point, i.e. for $\tilde{F} = \tilde{F}_c$, the relaxation time $-1/Re[\lambda]$ is typically denoted as the tunneling time $\tau$ in the context of optical bistability \cite{PhysRevA.35.1729}. In Fig. \ref{Fig3} (c) the dependence of $\tau$ on $N$ is presented together with a fit to an exponential decay. Note that for a first order phase transition at equilibrium the energy gap closes exponentially as a function of the size of the system. We conclude that in the thermodynamic limit the tunneling time diverges corresponding to a closing of the Liouvilian gap, i.e. $\lim_{N \to \infty}\lambda(\tilde{F} = \tilde{F}_c) \rightarrow 0$. 
If $\lambda = 0$ there are two eigenvectors of the Liouvillian with eigenvalue zero corresponding to two steady-state density matrices. This agrees with the semiclassical prediction of bistability.

As $\tilde{F} - \tilde{F}_c$ is decreased for finite $N$ the relaxation time $-1/Re[\lambda]$ exhibits a power law behavior and eventually converges to the tunneling time (see Fig. \ref{Fig3} (a)). This behavior is similar to a phase transition for a system at equilibrium whose energy gap typically closes according to a power law with a critical exponent \cite{book:577981}. However, in stark contrast we find that in the present case the power law exponent depends on the parameter $N$ and diverges in the thermodynamic limit. In order to quantify this we have fitted the power laws observed in Fig. \ref{Fig3} (a) as
\begin{eqnarray}
-1/Re[\lambda] = [(\tilde{F}-\tilde{F}_c)/f]^{-bN}/\gamma,
\label{Fit}
\end{eqnarray}
with $b$ and $f$ two fitting parameters (see the dash-dotted lines in Fig. \ref{Fig3} (a)). This parametrization of the fitting curve (\ref{Fit}) shows that for $N \rightarrow \infty$ the power law becomes a vertical curve at $\tilde{F} - \tilde{F}_c = f$. The obtained fitting parameters are plotted in Fig. \ref{Fig3} (d) and (e) as a function of $N$. In the thermodynamic limit they converge to $\lim_{N \to \infty}b \simeq 0.35$ and $\lim_{N \to \infty}f \simeq 0.214\gamma$ (these values are determined from a finite size scaling as demonstrated in the inset of Fig \ref{Fig3} (e) for $f$). This value for $f$ is in good agreement with $\tilde{F}^+-\lim_{N \to \infty}\tilde{F}_c \simeq 0.24\gamma$ (indicated by the dashed lines in Fig. \ref{Fig3} (a) and (e)), with $\tilde{F}^+= 1.16$, the value where the semiclassical edge of bistability occurs \cite{0305-4470-13-2-034}.
 The difference between the fitted values $0.214$ and $0.24$ is due to statistical error propagation in the two sequential fitting procedures \footnote{Each point of Fig. \ref{Fig3} (e) is obtained by fitting the data in Fig. \ref{Fig3} (a). By analyzing carefully our data, we estimate a standard deviation of 0.01-0.02 for each point in Fig.  \ref{Fig3} (e), which then propagates in the fit for the value of $f$ in the limit of large $N$. Hence, the small discrepancy is within the error bars of our fitting procedure.}.
In the thermodynamic limit there is a finite region where the Liouvillian gap closes, corresponding to two degenerate steady-state density matrices and resulting in optical bistability. Note that the same exponential vanishing of the Liouvillian gap is found by changing the nonlinearity and the detuning within the optical bistability regime, but the values of the coefficients $b$ and $f$ do depend on the parameters $U$ and $\Delta$. 

\section{0D Thermodynamic limit and link with driven-dissipative Bose-Hubbard model \label{TL}}
Let us now make a link with some other works where a similar thermodynamic limit in zero dimensions was considered.
In the context of Carmichael's discussion in Ref. \cite{PhysRevX.5.031028} we have considered a "weak-coupling thermodynamic limit" since the interaction strength $U = \tilde{U}/N$ goes to zero. In qualitative agreement with our results the numerical study of Carmichael in Ref. \cite{PhysRevX.5.031028} for the steady-state suggest a convergence to the semiclassical prediction in the regime with many photons. Recently the thermodynamic limit was also introduced to examine a second order phase transition and the associated entanglement of the driven-dissipative Bose-Hubbard dimer \cite{2016arXiv160702578C}. Such a thermodynamic limit has also been explored for the study of the cavity-QED laser threshold \cite{PhysRevA.50.4318} and for the conservative (no dissipation) Rabi and Jaynes-Cummings models \cite{PhysRevLett.115.180404, PhysRevLett.117.123602}. 

The considered thermodynamic limit of large photon numbers for one single-mode nonlinear resonator has been obtained while keeping constant the nonlinearity-intensity product $U F^2$: this might seem  artificial at first sight. Here we show that there is an intriguing connection with the traditional thermodynamical limit of many resonators in the driven-dissipative Bose-Hubbard model. Let us consider the Bose-Hubbard Hamiltonian with a homogeneous coherent drive given by 
\begin{eqnarray}
\hat{H}_{BH} =  &&-J\sum_{<i,j>}\left(\hat{a}_i^\dagger\hat{a}_j+ h.c.\right) + \sum_{i}\left( \tilde{F}e^{-i\omega_pt}\hat{a}_i^\dagger+h.c. \right)
 \nonumber \\ 
&&+ \sum_{i}\left(\omega_c\hat{a}_i^\dagger\hat{a}_i+\frac{\tilde{U}}{2}\hat{a}_i^\dagger\hat{a}_i^\dagger\hat{a}_i\hat{a}_i\right),   
\label{HamBH}
\end{eqnarray}  
with $J$ the hopping parameter, $\tilde{U}$ is the photon-photon interaction strength, $\tilde{F}$ the amplitude of the coherent drive with frequency $\omega_p$ and $\omega_c$ is the mode frequency of the resonators. An alternative and equivalent description is obtained by Fourier transforming from the real space to the dual reciprocal $\bm{k}$-space. The annihilation operator for an excitation in mode $\bm{k}$ is $\hat{a}_{\bm{k}} = 1/\sqrt{V}\sum_ie^{-i\bm{k}.\bm{r}}\hat{a}_i$, with $V$ the total volume and $\hat{a}_i$ the annihilation operator in the position space $\bm{r}$. Since the drive is homogeneous only the $\bm{k} = \bm{0}$ mode is externally driven and the other $\bm{k} \neq \bm{0}$ modes can get populated only through nonlinear scattering. Neglecting these non-homogeneous $\bm{k} \neq \bm{0}$ modes results in the following Hamiltonian for the $\bm{k} = \bm{0}$ mode:   
\begin{equation}
\hat{H}_{ \bm{0}} =  \omega_0 \hat{a}_{\bm{0}}^\dagger\hat{a}_{\bm{0}} + \frac{\tilde{U}}{2N} \hat{a}_{\bm{0}}^\dagger\hat{a}_{\bm{0}}^\dagger\hat{a}_{\bm{0}}\hat{a}_{\bm{0}}  
+ \sqrt{N} \left( \tilde{F}e^{-i\omega_pt}\hat{a}_{\bm{0}}^\dagger+ h.c.  \right),
\label{Ham}
\end{equation}  
with $\omega_0$ the $\bm{k} = \bm{0}$ mode frequency and $N$ the number of cavities. The dissipation does not couple the different $\bm{k}$-modes and has the same form in $\bm{k}$-space as in real space. By neglecting the $\bm{k} \neq \bm{0}$ modes the homogeneous driven-dissipative Bose-Hubbard model can thus be mapped exactly onto the driven-dissipative Kerr model considered in this letter. The $\bm{k} = \bm{0}$ mode frequency $\omega_0$ depends on the lattice geometry and the hopping parameter $J$. The effective nonlinearity $\tilde{U}/N$ and drive amplitude $\sqrt{N}\tilde{F}$ exhibit the previously considered scaling behavior with the parameter $N$ which now has a clear physical role as the number of cavities. This establishes a clear link between the thermodynamic limit we consider for a single mode and the more commonly considered thermodynamic limit of an infinite system size. 
%We note that in Ref. \cite{2016arXiv160106857W} it was found that for a 1D lattice of hard core bosons the Liouvillian gap decreases for a larger lattice which is in qualitative agreement with the interpretation of $N$ as the number of cavities. {\color{blue} What happens for soft core bosons remains to be explored.
%In the case of lattice of nonlinear resonators, 
The validity of neglecting the non-homogeneous $\bm{k} \neq \bm{0}$ modes is however not obvious, especially for a critical system and opens an intriguing perspective. In particular, as it happens for some equilibrium phase transitions, in one-dimensional arrays where correlations are more important the behavior could be significantly different from two-dimensional lattices and the vanishing of the Liouvillian gap in the thermodynamical limit is not guaranteed.

\section{Conclusions \label{Con}}
We have theoretically explored the closing of the Liouvillian spectral gap in a well defined thermodynamical limit of large excitation numbers for a driven-dissipative Kerr nonlinear resonator. Our work provides a clear paradigm of critical dynamical behavior for a dissipative first-order phase transition as it will stimulate further systematic studies of the Liouvillian gap and its finite size scaling for other critical dissipative model systems. It paves the way to intriguing experimental investigations in a broad class of platforms where Kerr optical nonlinearities can be implemented in single-mode resonators and in more complex systems such as lattices of coupled resonators.

\acknowledgements{We acknowledge fruitfull discussions with N. Bartolo and F. Minganti. W. C. and C. C. acknowledge support from ERC (via the Consolidator Grant "CORPHO" No. 616233). R. F. acknowledges support from EU-IP-SIQS, EU-IP-QUIC, SNS 2014 Projects, QSYNC. We are also grateful to the Kavli Institute for Theoretical Physics for their hospitality is the context of the program "Many-Body Physics with Light". }

\bibliography{manusc}

\end{document}